\documentclass[11pt]{article}
\usepackage{graphicx}
\usepackage{cite}

\begin{document}

%
%
\title{\Large\bf 
 Momentum sum rules for fragmentation functions}

\author{S.~Meissner$^{1}$, A.~Metz$^{2}$, and D.~Pitonyak$^{2}$
 \\[0.3cm]
{\normalsize\it $^1$Institut f{\"u}r Theoretische Physik II,} 
{\normalsize\it Ruhr-Universit{\"a}t Bochum, 44780 Bochum, Germany} \\[0.1cm]
{\normalsize\it $^2$Department of Physics, Barton Hall,} 
{\normalsize\it Temple University, Philadelphia, PA 19122-6082, USA}}

\maketitle


%
%
\begin{abstract}
\noindent
Momentum sum rules for fragmentation functions are considered.
In particular, we give a general proof of the Sch\"afer-Teryaev sum rule for 
the transverse momentum dependent Collins function.
We also argue that corresponding sum rules for related fragmentation functions 
do not exist.
Our model-independent analysis is supplemented by calculations in a simple 
field-theoretical model.
\end{abstract}

%
%
\section{Introduction}
\noindent
Fragmentation functions (FFs) contain important information about strong 
interaction dynamics in the non-perturbative regime.
It turns out that a realistic modeling of FFs is nontrivial.
Moreover, as a matter of principle, FFs cannot be computed in lattice gauge
theory.
In this situation, it is desirable to obtain as many model-independent constraints 
on these objects as possible.
Momentum sum rules do provide such constraints, with the momentum sum rule for 
the (collinear) unpolarized fragmentation function $D_1$ representing the best 
known example~\cite{Collins:1981uw}.
Any phenomenological parameterization of $D_1$ must obey this sum 
rule~\cite{Kretzer:2000yf,Bourhis:2000gs,Kniehl:2000fe,Albino:2005me,Hirai:2007cx,deFlorian:2007aj,deFlorian:2007hc,Albino:2008fy}.
Intuitively, the $D_1$ sum rule follows from conservation of the longitudinal 
momentum of the fragmenting parton.
Though intuitive, a rigorous proof in QCD is nontrivial~\cite{Collins:1981uw}, 
and we also address this issue in the present note.

In recent years, there has been an increased interest in transverse momentum
dependent FFs, which not only contain information on the longitudinal momentum of 
the final state hadron but also on its transverse motion relative to the parton 
(see, e.g., Refs.~\cite{Collins:1992kk,Mulders:1995dh,Barone:2001sp,Bacchetta:2006tn,D'Alesio:2007jt}).
In this context, the Collins fragmentation function $H_1^\perp$~\cite{Collins:1992kk},
which describes the fragmentation of a transversely polarized quark into an unpolarized
hadron, plays an important role.
It belongs to the class of (naive) time-reversal odd (T-odd) FFs, which implies that 
it is nonzero only if there exists a nontrivial phase for the decay 
$q^{\ast} \to h X$ of the (virtual) quark into a hadron. 
Model calculations of $H_1^\perp$ can be found in 
Refs.~\cite{Artru:1995bh,Bacchetta:2001di,Bacchetta:2002tk,Gamberg:2003eg,Bacchetta:2003xn,Amrath:2005gv,Bacchetta:2007wc,Artru:2010st}.

The Collins function is of particular interest since it can serve as a tool 
for addressing the transversity parton distribution in semi-inclusive deep-inelastic
scattering (DIS)~\cite{Collins:1992kk}.
The relevant observable --- the so-called Collins asymmetry --- has already been 
measured by the HERMES and COMPASS Collaborations for a proton and a deuteron 
target~\cite{Airapetian:2004tw,Alexakhin:2005iw,Ageev:2006da,Diefenthaler:2007rj,Levorato:2008tv}.
In the case of a proton target, clearly nonzero effects have been 
found~\cite{Airapetian:2004tw,Diefenthaler:2007rj,Levorato:2008tv}.
Information about the Collins function is also available through a particular 
azimuthal asymmetry in $e^+ e^- \to h_1 h_2 X$~\cite{Boer:1997mf,Boer:1997qn,Boer:2008fr} 
for which data from the Belle Collaboration exist~\cite{Abe:2005zx,Seidl:2008xc}.
Analyses of the data on the Collins asymmetry in semi-inclusive DIS and on the 
azimuthal asymmetry in $e^+ e^- \to h_1 h_2 X$ not only provided information about 
the Collins 
function~\cite{Vogelsang:2005cs,Efremov:2006qm,Anselmino:2007fs,Anselmino:2008jk} 
but also about the transversity distribution~\cite{Anselmino:2007fs,Anselmino:2008jk},
which represented a milestone in transverse spin physics.

The primary purpose of our paper is to address the so-called Sch\"afer-Teryaev 
sum rule (ST sum rule) for the Collins function~\cite{Schafer:1999kn}.
This sum rule states that a particular moment of $H_1^{\perp}$ for a 
fragmenting quark vanishes when summing over all final state hadrons.
It was obtained on the basis of intuitive arguments about conservation of 
transverse momentum in the fragmentation process~\cite{Schafer:1999kn},
yet a general proof of the ST sum rule in QCD did not exist.
Here we provide such a proof and also argue that related transverse momentum 
dependent FFs do not obey a corresponding sum rule, which is at variance
with some statements in the 
literature~\cite{Schafer:1999kn,Boer:1999ya,Anselmino:2000vs}.
In addition to the model-independent analysis, we compute the relevant FFs in a 
simple self-consistent quark-pion coupling model, and this study confirms the 
model-independent results.

%
%
\section{Derivation of sum rules}
\noindent
In order to provide a model-independent derivation of the momentum sum rules for FFs, 
we start from the basic correlator defining the fragmentation of a quark into a single 
hadron~\cite{Collins:1981uw,Mulders:1995dh,Boer:2003cm,Bacchetta:2006tn}\footnote{Note 
that, in general, integrals for which no integration limits are written explicitly 
run from $-\infty$ to $+\infty$.},
\begin{eqnarray} \label{e:corr_1}
 \Delta^{[\Gamma]}(z,\vec{k}_T,S_h) & = & 
 \frac{1}{4z} \sum_{X} \int \frac{d\xi^+ \, d^2\vec{\xi}_T}{(2\pi)^3}
 \, e^{i k \cdot \xi}  
\nonumber \\ 
 & & \mbox{} \times \textrm{Tr}
 \Big[ \langle 0 | {\cal W}_{1}(\infty,\xi) \, \psi(\xi) | P_h,S_h;X \rangle
 \langle P_h,S_h;X | \bar{\psi}(0) \, {\cal W}_{2}(0,\infty) | 0 \rangle \,
 \Gamma \Big]_{\xi^- = 0} \,.
 \hphantom{aaa}    
\end{eqnarray}
In this definition a color-average for the fragmenting quark is implicit,
a flavor index is suppressed, and the trace acts in Dirac space with 
$\Gamma$ representing a Dirac matrix.
The final state hadron is specified through its 4-momentum $P_h$ and the
covariant spin vector $S_h$, which satisfy $P_h^2 = M_h^2$, $S_h^2 = -1$, and
$P_h \cdot S_h = 0$.
The correlator~(\ref{e:corr_1}) is understood in a frame in which the transverse 
momentum of the hadron vanishes, while $\vec{k}_T$ is the transverse momentum 
of the quark.
The (large) minus-component of the hadron 
momentum\footnote{For a generic 4-vector $v$, we define light-cone 
coordinates according to $v^{\pm} = (v^0 \pm v^3)/\sqrt{2}$ and
$\vec{v}_T = (v^1 , v^2)$.}
is given by $P_h^- = z k^-$.
Color gauge invariance is ensured by means of the two Wilson lines
\begin{eqnarray} \label{e:wilson_1}
{\cal W}_{1}(\infty,\xi) & = &
{\cal W}(\infty^+, 0^-, \vec{\infty}_T ; \infty^+, 0^-, \vec{\xi}_T) \,
{\cal W}(\infty^+, 0^-, \vec{\xi}_T ; \xi^+, 0^-, \vec{\xi}_T) \,,
\vphantom{\Big(} \\ \label{e:wilson_2}
{\cal W}_{2}(0,\infty) & = &
{\cal W}(0^+, 0^-, \vec{0}_T ; \infty^+, 0^-, \vec{0}_T) \,
{\cal W}(\infty^+, 0^-, \vec{0}_T ; \infty^+, 0^-, \vec{\infty}_T) \,,
\end{eqnarray}
where, in general, ${\cal W}(a^+, a^-, \vec{a}_T ; b^+, b^-, \vec{b}_T)$ 
indicates a Wilson line running from $(a^+, a^-, \vec{a}_T)$ to 
$(b^+, b^-, \vec{b}_T)$.
In connection with $k_T$-dependent parton correlators, the importance of 
transversely running gauge links at the light-cone infinity, like the ones showing 
up in~(\ref{e:wilson_1}) and (\ref{e:wilson_2}), has been pointed out only relatively 
recently~\cite{Ji:2002aa,Belitsky:2002sm,Boer:2003cm}.
These links do not disappear in the light-cone gauge $A^- = 0$.
Nevertheless, as we will argue, their presence does not spoil the
longitudinal momentum sum rule for $D_1$.
We also note that the path of the Wilson lines for transverse momentum 
dependent FFs is not entirely unique.
Information on this topic can be found in various 
articles~\cite{Metz:2002iz,Collins:2004nx,Yuan:2007nd,Yuan:2008yv,Gamberg:2008yt,Ma:2008cj,Meissner:2008yf,Yuan:2009dw}.
Our derivation of the momentum sum rules goes through for any allowed path.

The (eight) leading twist transverse momentum dependent FFs (for fragmentation
of a quark $q$ into a spin-$\frac{1}{2}$ hadron $h$) are defined through the 
correlator in~(\ref{e:corr_1}) according to~\cite{Mulders:1995dh}
\begin{eqnarray} \label{e:ff_1}
 \Delta^{[\gamma^-]}(z,\vec{k}_T,S_h) & = &
 D_1^{h/q}(z,z^2\vec{k}_T^2) + 
 \frac{\epsilon_T^{ij} k_T^i S_{hT}^j}{M_h} \, D_{1T}^{\perp \, h/q}(z,z^2\vec{k}_T^2) \,,
\\ \label{e:ff_2}
 \Delta^{[\gamma^- \gamma_5]}(z,\vec{k}_T,S_h) & = &
 \lambda_h \, G_{1L}^{h/q}(z,z^2\vec{k}_T^2) +
 \frac{\vec{k}_T \cdot \vec{S}_{hT}}{M_h} \, G_{1T}^{h/q}(z,z^2\vec{k}_T^2) \,,  
\\ \label{e:ff_3}
 \Delta^{[i\sigma^{i-} \gamma_5]}(z,\vec{k}_T,S_h) & = &
 S_{hT}^i \bigg( H_{1T}^{h/q}(z,z^2\vec{k}_T^2) + 
 \frac{\vec{k}_T^2}{2 M_h^2} \, H_{1T}^{\perp \, h/q}(z,z^2\vec{k}_T^2) \bigg)
 - \frac{\epsilon_{T}^{ij} k_T^j}{M_h} \, H_{1}^{\perp \, h/q}(z,z^2\vec{k}_T^2) 
\nonumber \\
& & 
 + \frac{\lambda_h k_T^i}{M_h} \, H_{1L}^{\perp \, h/q}(z,z^2\vec{k}_T^2)
 + \frac{2 k_T^i \, \vec{k_T} \cdot \vec{S}_{hT} - S_{hT}^i \vec{k}_T^2}{2 M_h^2} \,
   H_{1T}^{\perp \, h/q}(z,z^2\vec{k}_T^2) \,.
\end{eqnarray}
In these definitions we use $\epsilon_T^{ij} = \epsilon^{-+ij}$ 
(with the convention $\epsilon_T^{12}=1$) and the representation
\begin{equation}
S_h = \Big( S_h^+, S_h^-, \vec{S}_{hT} \Big) = 
\bigg(-\lambda_h \frac{M_h}{2 P_h^-}, \lambda_h \frac{P_h^-}{M_h}, \vec{S}_{hT} \bigg)
\end{equation}
of the covariant spin vector.

It is now convenient to switch to a reference frame in which the fragmenting quark 
has no transverse momentum.
This implies a nonzero transverse momentum of the hadron, and, if one wants to keep 
the minus-component of 4-momenta fixed, this transverse momentum is given by
$\vec{P}_{h\perp} = - z \vec{k}_T$~\cite{Collins:1981uw}.
One can therefore write the correlator in~(\ref{e:corr_1}) as
(see also, e.g., Ref.~\cite{Collins:1981uw})
\begin{eqnarray} \label{e:corr_2}
 \Delta^{[\Gamma]}(z,\vec{P}_{h\perp},S_h) & = & 
 \frac{1}{4z} \int \frac{d\xi^+ \, d^2\vec{\xi}_T}{(2\pi)^3}
 \, e^{i k^- \xi^+}  
\nonumber \\ 
 & & \mbox{} \times \textrm{Tr} 
 \Big[ \langle 0 | {\cal W}_{1}(\infty,\xi) \, \psi(\xi) \, 
 \hat{a}_h^{\dagger}(P_h,S_h) \, \hat{a}_h(P_h,S_h) \, 
 \bar{\psi}(0) \, {\cal W}_{2}(0,\infty) | 0 \rangle \,
 \Gamma \Big]_{\xi^- = 0} \,,
\end{eqnarray}
where we have expressed the final state hadron through the particle creation 
operator $\hat{a}_h^{\dagger}(P_h,S_h)$.
This leads to
\begin{eqnarray} \label{e:corr_3}
 \lefteqn{\sum_{S_h} \int_0^1 dz \int d^2\vec{P}_{h\perp} \, P_h^{\mu} \, 
          \Delta^{[\Gamma]}(z,\vec{P}_{h\perp},S_h)}
\nonumber \\ 
& & = \frac{1}{2} \, \int d\xi^+ \, d^2\vec{\xi}_T
 \, e^{i k^- \xi^+} \, \textrm{Tr}
 \Big[ \langle 0 | {\cal W}_{1}(\infty,\xi) \, \psi(\xi) \, \hat{P}_h^{\mu} \,
 \bar{\psi}(0) \, {\cal W}_{2}(0,\infty) | 0 \rangle \,
 \Gamma \Big]_{\xi^- = 0} \,,
\end{eqnarray}
with the momentum operator (in light-cone quantization)~\cite{Collins:1981uw}
\begin{equation} \label{e:momop_1}
 \hat{P}_h^{\mu} = \sum_{S_h} \int 
 \frac{d P_h^- \, d^2 \vec{P}_{h\perp}}{(2\pi)^3 \, 2P_h^-} \,
 \hat{a}_h^{\dagger}(P_h,S_h) \, P_h^{\mu} \, \hat{a}_h(P_h,S_h) \,.
\end{equation}
By summing over all hadrons $h$, we obtain the momentum operator of the theory
expressed through hadronic field operators~\cite{Collins:1981uw},
\begin{equation} \label{e:momop_2}
\sum_h \hat{P}_h^{\mu} = \hat{P}^{\mu} \,. 
\end{equation}
We emphasize that, according to~(\ref{e:momop_1}), the relation~(\ref{e:momop_2}) 
also involves a summation over particle spins.
As we discuss below in a bit more detail, this is the main reason why momentum sum 
rules for FFs describing hadron polarization do not exist.
Using the properties of the momentum operator in~(\ref{e:momop_2}) one finds
\begin{eqnarray} \label{e:corr_4}
 \lefteqn{\sum_h \sum_{S_h} \int_0^1 dz \int d^2\vec{P}_{h\perp} \, P_h^{\mu} \, 
          \Delta^{[\Gamma]}(z,\vec{P}_{h\perp},S_h)}
\nonumber \\ 
& & = \frac{1}{2} \, \int d\xi^+ \, d^2\vec{\xi}_T
 \, e^{i k^- \xi^+} \, i \partial^{\mu} \,
 \bigg[ \textrm{Tr}
 \Big[ \langle 0 | {\cal W}_{1}(\infty,\xi) \, \psi(\xi) \,
 \bar{\psi}(0) \, {\cal W}_{2}(0,\infty) | 0 \rangle \,
 \Gamma \Big] \bigg]_{\xi^- = 0} \,,
\end{eqnarray}
where we exploited the identity
\begin{equation}
 \langle 0 | {\cal W}_{1}(\infty,\xi) \, \psi(\xi) \, \hat{P}^{\mu} =
 i \partial^{\mu} \, \Big[ \langle 0 | {\cal W}_{1}(\infty,\xi) \, \psi(\xi) \Big] \,.
\end{equation}
On the basis of~(\ref{e:corr_4}) one can now derive both the momentum sum rule for 
$D_1$ as well as the ST sum rule for the Collins function.

We begin with the sum rule for $D_1$.
The essential elements of a complete proof of this sum rule in QCD were already 
indicated in~\cite{Collins:1981uw}.
(For a proper treatment of ultraviolet divergences in $k_T$-integrated FFs we also 
refer to~\cite{Collins:1981uw}.)
For completeness, and also because of the potential complications arising from the 
transversely running gauge links in~(\ref{e:wilson_1}) and (\ref{e:wilson_2}), we 
consider it worthwhile to write out some details of a proof in the light-cone 
gauge $A^- = 0$.
To this end we choose $\mu = -$ in Eq.~(\ref{e:corr_4}) and use integration by 
parts, leading to
\begin{eqnarray} \label{e:corr_5}
 \lefteqn{\sum_h \sum_{S_h} \int_0^1 dz \int d^2\vec{P}_{h\perp} \, P_h^{-} \, 
          \Delta^{[\Gamma]}(z,\vec{P}_{h\perp},S_h)}
\nonumber \\ 
& & = \frac{k^-}{2} \, 
 \int d\xi^+ \, d^2\vec{\xi}_T
 \, e^{i k^- \xi^+} \, 
 \textrm{Tr}
 \Big[ \langle 0 | {\cal W}_{1}(\infty,\xi) \, \psi(\xi) \,
 \bar{\psi}(0) \, {\cal W}_{2}(0,\infty) | 0 \rangle \,
 \Gamma \Big]_{\xi^- = 0} \,.
\end{eqnarray}
Now we consider~(\ref{e:corr_5}) for $\Gamma = \gamma^-$ and introduce the 
so-called ``good'' quark field 
$\psi_{-} = \frac{1}{2} \gamma^+ \gamma^- \psi$~\cite{Kogut:1969xa,Jaffe:1996zw} 
providing
\begin{eqnarray} \label{e:corr_6}
 \lefteqn{\sum_h \sum_{S_h} \int_0^1 dz \int d^2\vec{P}_{h\perp} \, P_h^{-} \, 
          \Delta^{[\gamma^-]}(z,\vec{P}_{h\perp},S_h)}
\nonumber \\ 
& & = \frac{k^-}{\sqrt{2}} \, 
 \int d\xi^+ \, d^2\vec{\xi}_T
 \, e^{i k^- \xi^+} \, 
 \textrm{Tr} \Big[ \langle 0 | {\cal W}_{1}(\infty,\xi) \, \psi_{-}(\xi) \,
 \psi_{-}^{\dagger}(0) \, {\cal W}_{2}(0,\infty) | 0 \rangle
 \Big]_{\xi^- = 0}
\nonumber \\
& & = \frac{k^-}{\sqrt{2}} \, 
 \int d\xi^+\, d^2\vec{\xi}_T
 \, e^{i k^- \xi^+} \, 
 \textrm{Tr} \Big[ \langle 0 | 
 {\cal W}(\infty^+, 0^-, \vec{\infty}_T ; \infty^+, 0^-, \vec{\xi}_T) 
\nonumber \\ 
& & \hspace{2.0cm} \mbox{} \times 
 \psi_{-}(\xi^+, 0^-, \vec{\xi}_T) \, \psi_{-}^{\dagger}(0^+, 0^-, \vec{0}_T) \, 
 {\cal W}(\infty^+, 0^-, \vec{0}_T ; \infty^+, 0^-, \vec{\infty}_T)  
 | 0 \rangle \Big]
\nonumber \\
& & = \frac{k^-}{\sqrt{2}} \, 
 \int d\xi^+ \, d^2\vec{\xi}_T
 \, e^{i k^- \xi^+} \, 
 \textrm{Tr} \Big[ \langle 0 | 
 {\cal W}(\infty^+, 0^-, \vec{\infty}_T ; \infty^+, 0^-, \vec{\xi}_T) 
\nonumber \\ 
& & \hspace{2.0cm} \mbox{} \times 
 \{ \psi_{-}(\xi^+, 0^-, \vec{\xi}_T) \, , \, \psi_{-}^{\dagger}(0^+, 0^-, \vec{0}_T) \} \, 
 {\cal W}(\infty^+, 0^-, \vec{0}_T ; \infty^+, 0^-, \vec{\infty}_T)  
 | 0 \rangle \Big] \,.
\end{eqnarray}
In the second step in~(\ref{e:corr_6}) we made use of the light-cone gauge 
$A^- = 0$, for which the Wilson lines in~(\ref{e:wilson_1}) and (\ref{e:wilson_2}) 
that run along the light-cone reduce to unity.
In the last step the anti-commutator of the two quark fields was introduced, which
is justified because of 
\begin{eqnarray} \label{e:corr_7}
& & \int d\xi^+\, d^2\vec{\xi}_T
 \, e^{i k^- \xi^+} \, 
 \textrm{Tr} \Big[ \langle 0 | 
 {\cal W}(\infty^+, 0^-, \vec{\infty}_T ; \infty^+, 0^-, \vec{\xi}_T) 
\nonumber \\ 
& & \hspace{2.0cm} \mbox{} \times 
 \psi_{-}^{\dagger}(0^+, 0^-, \vec{0}_T) \, \psi_{-}(\xi^+, 0^-, \vec{\xi}_T) \, 
 {\cal W}(\infty^+, 0^-, \vec{0}_T ; \infty^+, 0^-, \vec{\infty}_T)  
 | 0 \rangle \Big]
\nonumber \\
& = & \sum_X \int d\xi^+\, d^2\vec{\xi}_T
 \, e^{i k^- \xi^+} \, e^{i \sum_j p_j^- \xi^+} \,
 \textrm{Tr} \Big[ \langle 0 | 
 {\cal W}(\infty^+, 0^-, \vec{\infty}_T ; \infty^+, 0^-, \vec{\xi}_T) 
\nonumber \\ 
& & \hspace{2.0cm} \mbox{} \times 
 \psi_{-}^{\dagger}(0^+, 0^-, \vec{0}_T) | X \rangle \langle X | 
 \psi_{-}(0^+, 0^-, \vec{\xi}_T) \, 
 {\cal W}(\infty^+, 0^-, \vec{0}_T ; \infty^+, 0^-, \vec{\infty}_T)  
 | 0 \rangle \Big]
\nonumber \\
& = & (2\pi) \sum_X \int d^2\vec{\xi}_T \,
 \delta \Big( k^- + \sum_j p_j^- \Big) \,
 \textrm{Tr} \Big[ \langle 0 | 
 {\cal W}(\infty^+, 0^-, \vec{\infty}_T ; \infty^+, 0^-, \vec{\xi}_T) 
\nonumber \\ 
& & \hspace{2.0cm} \mbox{} \times 
 \psi_{-}^{\dagger}(0^+, 0^-, \vec{0}_T) | X \rangle \langle X | 
 \psi_{-}(0^+, 0^-, \vec{\xi}_T) \, 
 {\cal W}(\infty^+, 0^-, \vec{0}_T ; \infty^+, 0^-, \vec{\infty}_T)  
 | 0 \rangle \Big]
\nonumber \\
& = & 0 \,. \vphantom{\Big(}
\end{eqnarray}
In Eq.~(\ref{e:corr_7}), $p_j$ are the 4-momenta of the particles in the 
intermediate states $| X \rangle$.
The expression vanishes since $k^- > 0$ and $p_j^- \ge 0$.

In order to proceed with~(\ref{e:corr_6}), one can use the anti-commutator 
for the ``good'' quark fields~\cite{Kogut:1969xa,Collins:1981uw},
\begin{equation}
\{ \psi_{-}(\xi^+, 0^-, \vec{\xi}_T) \, , \, \psi_{-}^{\dagger}(0^+, 0^-, \vec{0}_T) \} 
= \frac{1}{2\sqrt{2}} \gamma^+ \gamma^- \delta(\xi^+) \delta^{(2)}(\vec{\xi}_T) \,,
\end{equation}
which immediately gives
\begin{equation} \label{e:sr1_h1}
 \sum_h \sum_{S_h} \int_0^1 dz \int d^2\vec{P}_{h\perp} \, P_h^{-} \, 
 \Delta^{[\gamma^-]}(z,\vec{P}_{h\perp},S_h)
 = k^- \,.
\end{equation}
On the other hand, because of~(\ref{e:ff_1}), one also has
\begin{equation} \label{e:sr1_h2}
 \sum_h \sum_{S_h} \int_0^1 dz \int d^2\vec{P}_{h\perp} \, P_h^{-} \, 
 \Delta^{[\gamma^-]}(z,\vec{P}_{h\perp},S_h)
 = 
 \sum_h \sum_{S_h} \int_0^1 dz \int d^2\vec{P}_{h\perp} \, z k^{-} \, 
 D_1^{h/q}(z,\vec{P}_{h\perp}^2) \,.
\end{equation}
Comparing Eqs.~(\ref{e:sr1_h1}) and~(\ref{e:sr1_h2}), and going back to the original 
reference frame in which $\vec{P}_{h\perp} = 0$, then leads to the momentum sum rule for 
$D_1$,
\begin{eqnarray} \label{e:sr1}
& & \sum_h \sum_{S_h} \int_0^1 dz \, z \, D_1^{h/q}(z) = 1 \,,
\\ \label{e:ff_mom1}
& & \textrm{with} \quad
D_1^{h/q}(z) = z^2 \int d^2 \vec{k}_T \, D_1^{h/q}(z,z^2\vec{k}_T^{\,2}) \,.
\end{eqnarray}
According to~(\ref{e:ff_1}), $D_1$ for a spin-$\frac{1}{2}$ hadron is 
defined by a spin average rather than a spin summation.
Therefore, in the sum rule a summation over hadron spins shows up, which implies 
that one has to multiply FFs for a spin-$\frac{1}{2}$ particle by 2.
Also note that a corresponding sum rule for the two collinear FFs
\begin{eqnarray} \label{e:coll_1}
G_1^{h/q}(z) & = & z^2 \int d^2 \vec{k}_T \, G_{1L}^{h/q}(z,z^2\vec{k}_T^{\,2}) \,,
\\ \label{e:coll_2}
H_1^{h/q}(z) & = & z^2 \int d^2 \vec{k}_T \, \bigg( H_{1T}^{h/q}(z,z^2\vec{k}_T^2) + 
 \frac{\vec{k}_T^2}{2 M_h^2} \, H_{1T}^{\perp \, h/q} (z,z^2\vec{k}_T^2) \bigg)
\end{eqnarray}
cannot be derived along the lines described above.
These functions drop out when summing the fragmentation correlators (\ref{e:ff_2})
and (\ref{e:ff_3}) over the hadron polarizations.
However, as we pointed out after~(\ref{e:momop_2}), this summation is a crucial 
element in the proof of momentum sum rules for FFs.
Since one also finds that for $G_1$ and $H_1$ the respective traces vanish, 
i.e., the right hand side of the formulas corresponding to (\ref{e:sr1_h1}) 
vanishes, one arrives at the consistent though useless situation $0=0$.

Now we turn to the (simpler) derivation of the ST sum rule.
Starting again from Eq.~(\ref{e:corr_4}) and choosing $\mu = j$, with $j$ being a 
transverse index, one readily finds
\begin{equation} \label{e:sr2_h1}
 \sum_h \sum_{S_h} \int_0^1 dz \int d^2\vec{P}_{h\perp} \, P_h^{j} \, 
 \Delta^{[\Gamma]}(z,\vec{P}_{h\perp},S_h) = 0 \,.
\end{equation}
This result holds because of
\begin{equation}
 \int d\xi^j \, \partial^j  \, {\cal W}_{1}(\infty,\xi) \, \psi(\xi) 
 = {\cal W}_{1}(\infty,\xi) \psi(\xi) \Big|_{\xi_T^j = \infty} - 
   {\cal W}_{1}(\infty,\xi) \psi(\xi) \Big|_{\xi_T^j = - \infty} \,,  
\end{equation}
and the vanishing of the quark field at $\xi_T^j = \pm \infty$.
On the other hand, from Eq.~(\ref{e:ff_3}) one obtains
\begin{eqnarray} 
 \lefteqn{\sum_h \sum_{S_h} \int_0^1 dz \int d^2\vec{P}_{h\perp} \, P_h^{j} \, 
 \Delta^{[i \sigma^{i-} \gamma_5]}(z,\vec{P}_{h\perp},S_h)}
\nonumber \\  \label{e:sr2_h2}
& & = \epsilon_T^{ij} \sum_h \sum_{S_h} \int_0^1 dz \int d^2\vec{P}_{h\perp} \, 
 \frac{\vec{P}_{h\perp}^2}{2 z M_h} \, H_1^{\perp \, h/q}(z,\vec{P}_{h\perp}^2) \,.
\end{eqnarray}
Comparing Eqs.~(\ref{e:sr2_h1}) and~(\ref{e:sr2_h2}), and going back to the original 
reference frame in which $\vec{P}_{h\perp} = 0$, then leads to the ST sum rule
for the Collins function~\cite{Schafer:1999kn} in the form
\begin{eqnarray} \label{e:sr2}
& & \sum_h \sum_{S_h} \int_0^1 dz \, z \, M_h \, H_{1}^{\perp(1) \, h/q}(z) = 0 \,,
\\ \label{e:ff_mom2}
& & \textrm{with} \quad
H_{1}^{\perp(1) \, h/q}(z) = z^2 \int d^2 \vec{k}_T \, 
 \frac{\vec{k}_T^{\,2}}{2 M_h^2} \, H_1^{\perp \, h/q}(z,z^2\vec{k}_T^{\,2}) \,.
\end{eqnarray}
Since we stick to the conventions of Ref.~\cite{Mulders:1995dh}, the hadron mass
$M_h$ appears in~(\ref{e:sr2}).
This factor would not show up if in~(\ref{e:ff_3}) and~(\ref{e:ff_mom2}) a common 
mass scale for each hadron was used.
As mentioned earlier, this sum rule was already obtained in 
Ref.~\cite{Schafer:1999kn} (with slightly different conventions) on the basis
of intuitive arguments about conservation of transverse momentum in the 
fragmentation process.
However, a general field-theoretical proof was not yet available.
In fact, the same argument about conservation of transverse momentum led to 
the conclusion that sum rules corresponding to the one in~(\ref{e:sr2}) should 
also hold for other transverse momentum dependent FFs~\cite{Schafer:1999kn}.
More precisely, sum rules of the type~(\ref{e:sr2}) were expected for
$D_{1T}^\perp$, $G_{1T}$, and $H_{1L}^\perp$ since in 
Eqs.~(\ref{e:ff_1})--(\ref{e:ff_3}) those FFs, like the Collins function, 
are accompanied by a term linear in $k_T$. 
Basically by repeating the reasoning we used above in connection with the 
collinear FFs $G_1$ and $H_1$ in~(\ref{e:coll_1}) and~(\ref{e:coll_2}), 
one finds that the proof of the ST sum rule cannot be extended to other transverse 
momentum dependent FFs.
One rather ends up again with the situation $0 = 0$.
Below we will explicitly show by model calculations that 
$D_{1T}^\perp$, $G_{1T}$, and $H_{1L}^\perp$ do not obey a sum rule like the
ST sum rule in~(\ref{e:sr2}).

%
%
\section{Model calculations}
\noindent
In this section we explore the momentum sum rules for the FFs in a simple though
self-consistent field-theoretical model.
To describe the matrix elements in the fragmentation correlator, we use a 
pseudoscalar coupling between quarks and pions given by the interaction Lagrangian
\begin{equation} \label{e:model}
{\mathcal L}_I(x) = - i g \, \bar{\psi}(x) \, \gamma_5 \, \psi(x) \, \pi(x) \,,
\end{equation}
which is in the spirit of the Manohar-Georgi model~\cite{Manohar:1983md}.
For simplicity we do not take a flavor degree of freedom into account, which
is sufficient for our purpose.
This model was already exploited in Ref.~\cite{Bacchetta:2001di} in order to get
an explicit realization of a nonzero Collins function for a pion.
A slightly modified/extended version of this model was also studied recently 
with the main aim of obtaining a reasonable phenomenology for $D_1^{\pi/q}$ by 
taking into account multiple pion emission~\cite{Ito:2009zc}.

\begin{figure}[t]
\begin{center}
\includegraphics[width=12cm]{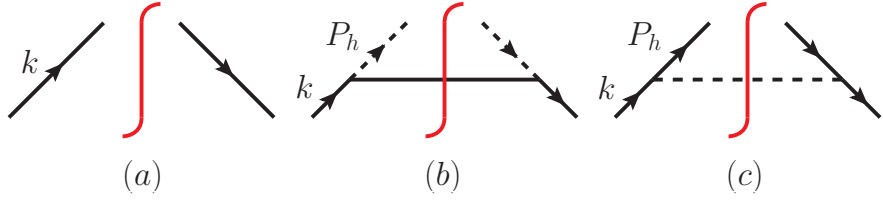}
\caption{Cut-diagrams describing the fragmentation of a quark (solid line) into
a quark or a pion (dashed line) through ${\cal O}(g^2)$ in a quark-pion coupling 
model as specified in Eq.~(\ref{e:model}).}
\label{f:1}
\end{center}
\end{figure}

\begin{figure}[t]
\begin{center}
\includegraphics[width=12cm]{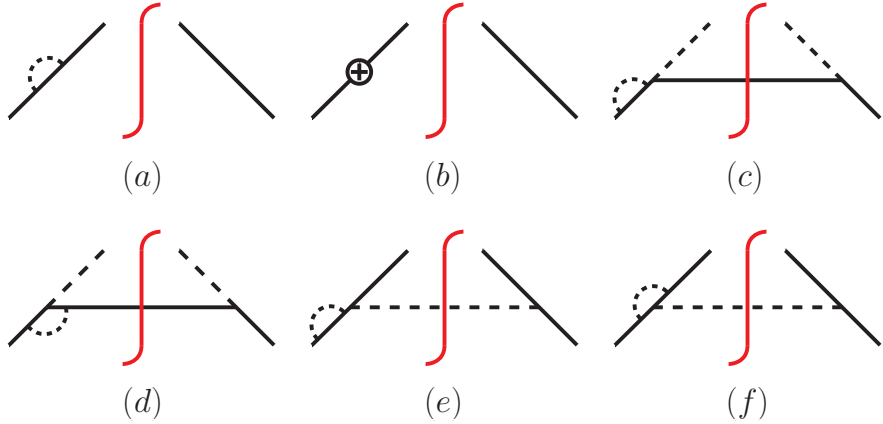}
\caption{Loop contributions describing the fragmentation of a quark (solid line) into
a quark or a pion (dashed line) in a quark-pion coupling model as specified in 
Eq.~(\ref{e:model}).
Hermitian conjugate graphs are not shown.
Diagram (a) is canceled when adding it to the respective counter term contribution 
in (b).
Diagrams (c)--(f) generate nonzero results for the T-odd FFs $H_1^\perp$ and 
$D_{1T}^\perp$.}
\label{f:2}
\end{center}
\end{figure}

In the case of $D_1$, we compute all the contributions through ${\cal O}(g^2)$.
In our model, the summation over all hadrons in the sum rule~(\ref{e:sr1}) implies
a summation both over pions and quarks in the final state.
One obtains (see also Refs.~\cite{Bacchetta:2001di,Ito:2009zc}) 
\begin{eqnarray} \label{e:d1_res1}
 D_1^{q/q}(z,z^2 \vec{k}_T^{\,2}) & = &
 \frac{1}{2} \, \delta(1-z) \, \delta^{(2)}(\vec{k}_T) \, Z_{\psi}
 + \frac{g^2}{32\pi^3} \,
 \frac{(1-z) \Big( \vec{k}_T^{\,2} + \frac{(1-z)^2}{z^2} m^2 \Big)}
     {z^2 \, \Big( \vec{k}_T^{\,2} + \frac{(1-z)^2}{z^2} m^2 
               + \frac{m_{\pi}^2}{z} \Big)^2} \,,
\\ \label{e:d1_res2}
 D_1^{\pi/q}(z,z^2 \vec{k}_T^{\,2}) & = &
 \frac{g^2}{16\pi^3} \,
 \frac{\vec{k}_T^{\,2} + m^2}
      {z \, \Big (\vec{k}_T^{\,2} + m^2 + \frac{1-z}{z^2}m_{\pi}^2 \Big)^2} \,,
\end{eqnarray}
with $m$ denoting the quark mass and $m_{\pi}$ the pion mass.
The first term in~(\ref{e:d1_res1}) arises from diagram (a) in 
Fig.~\ref{f:1}, which represents the lowest order contribution from a vacuum 
intermediate state.
Note that this term must also include the wave function renormalization 
factor~\cite{Collins:1981uw,Ito:2009zc}
\begin{equation}
 Z_{\psi} = 1 + \frac{\partial \Sigma}
                     {\partial k \hspace{-0.20cm}/ \,} \Big|_{k \hspace{-0.17cm} / \, = m} \,,
\end{equation}
which in our case is given by the quark self energy $\Sigma$ to one loop.
The second term in~(\ref{e:d1_res1}) describes the contribution from diagram (c), 
while diagram (b) in Fig.~\ref{f:1} leads to the result in Eq.~(\ref{e:d1_res2}).
A potential contribution to $D_1^{q/q}$ at ${\cal O}(g^2)$ from diagram (a) in 
Fig.~\ref{f:2} is canceled by the counter term diagram (b).

Now we consider the momentum sum rule~(\ref{e:sr1}), for which we find
\begin{eqnarray} \label{e:d1_sr_mod}
\lefteqn{\int_0^1 dz \, z \, \Big( 2 \, D_1^{q/q}(z) + D_1^{\pi/q}(z) \Big)}  
\nonumber \\
& & = 1 + \frac{g^2}{16\pi^3} \int_0^1 dz \int d^2 \vec{l}_T \,
 \frac{\vec{l}_T^{\; 2} (2z-1) + m^2 z^2 - m_{\pi}^2 (1-z)^2}
      {\Big(\vec{l}_T^{\; 2} + m^2 z^2 + m_{\pi}^2 (1-z) \Big)^2}
\nonumber \\
& & = 1 + \frac{g^2}{16\pi^2} \int_0^1 dz 
 \bigg( (1-2z) \ln(z^2 + \mu^2 (1-z)) + 
         \frac{z^2 - \mu^2 (1-z)^2}{z^2 + \mu^2 (1-z)} \bigg)
\nonumber \\
& & = 1 \vphantom{\Big(} \,,
\end{eqnarray}
i.e., the sum rule holds in the model~(\ref{e:model}) through ${\cal O}(g^2)$.
Note that we introduced the mass ratio $\mu = m_{\pi}/m$.
In the case of the contribution from $Z_{\psi}$ in the second line 
of~(\ref{e:d1_sr_mod}), $z$ is a Feynman parameter and $l_T$ is the transverse 
part of the loop momentum.
To carry out the ultraviolet divergent $l_T$-integral, one can use dimensional
regularization or a cutoff, with both methods leading to the third line 
in~(\ref{e:d1_sr_mod}).
The vanishing of the remaining $z$-integral is an exact analytical result.
As an independent check, we have computed all the contributions right from 
the beginning in $4-\epsilon$ dimensions. 
Then the sum rule can also be verified if one keeps in mind 
that~\cite{Collins:1981uw}
\begin{equation}
D_1(z) = 
\int d^{2 - \epsilon} P_{h\perp} \, D_1(z,\vec{P}_{h\perp}^2) =
z^{2 - \epsilon} \int d^{2 - \epsilon} k_T \, D_1(z,z^2\vec{k}_T^2) \,.
\end{equation}
In particular, the non-integral exponent in the factor $z^{2 - \epsilon}$ is crucial 
for getting the desired result.
It is perhaps worth mentioning that for either way we had to carry out all 
the integrations to the very end in order to establish the momentum sum rule 
in~(\ref{e:sr1}).
In this respect our discussion of the $D_1$ sum rule differs from the 
corresponding one given in~\cite{Ito:2009zc}.

Next, we turn our attention to the Collins function and the ST sum rule in
Eq.~(\ref{e:sr2}).
The Collins function receives contributions from diagrams (c)--(f) in 
Fig.~\ref{f:2}, and after some algebra one obtains\footnote{The result for
$H_1^{\perp \, \pi/q}$ was already given in~\cite{Bacchetta:2001di}, but the 
overall sign was wrong as pointed out previously in Ref.~\cite{Bacchetta:2003xn}.} 
\begin{eqnarray} \label{e:coll_res1}
H_1^{\perp \, q/q}(z,z^2 \vec{k}_T^2) & = &
\frac{g^2}{16 \pi^3} \frac{m}{1-z} 
\bigg( \frac{m \, \textrm{Im} \, \tilde{\Sigma}(k^2)}{(k^2 - m^2)^2} 
     + \frac{\textrm{Im} \, \tilde{\Gamma}_{q}(k^2)}{k^2 - m^2} \bigg) 
       \bigg|_{k^2 = \frac{z}{1-z}\vec{k}_T^2 + \frac{m^2}{z} + \frac{m_{\pi}^2}{1-z}} \,,
\\ \label{e:coll_res2}
H_1^{\perp \, \pi/q}(z,z^2 \vec{k}_T^2) & = &
- \frac{g^2}{8 \pi^3} \frac{m_{\pi}}{1-z} 
\bigg( \frac{m \, \textrm{Im} \, \tilde{\Sigma}(k^2)}{(k^2 - m^2)^2} 
     + \frac{\textrm{Im} \, \tilde{\Gamma}_{\pi}(k^2)}{k^2 - m^2} \bigg) 
       \bigg|_{k^2 = \frac{z}{1-z}\vec{k}_T^2 + \frac{m^2}{1-z} + \frac{m_{\pi}^2}{z}} \,.
\end{eqnarray}
In these expressions $\textrm{Im}\,\tilde{\Sigma}$ arises from the self energy 
insertions in the diagrams (c) and (e) in Fig.~\ref{f:2}, while 
$\textrm{Im}\,\tilde{\Gamma}_{q}$ and $\textrm{Im}\,\tilde{\Gamma}_{\pi}$, 
respectively, are due to the vertex corrections in the diagram (f) and (d).
Note that for the expressions in Eq.~(\ref{e:coll_res1}) and~(\ref{e:coll_res2}) 
the virtuality $k^2$ of the fragmenting quark has a different value.
If one actually evaluates them at the same $k^2$, one can show that
\begin{equation} \label{e:vertex_id}
\textrm{Im}\,\tilde{\Gamma}_{\pi}(k^2) = \textrm{Im}\,\tilde{\Gamma}_{q}(k^2) \,,
\end{equation}
which is quite essential for verifying the ST sum rule.
Though the explicit results of the imaginary parts turn out to be irrelevant for 
the discussion of the ST sum rule, we include them here for 
completeness~\cite{Bacchetta:2001di}:
\begin{eqnarray}
\textrm{Im} \, \tilde{\Sigma}(k^2) & = & \frac{g^2}{16 \pi^2} 
\bigg( 1 - \frac{m^2 - m_{\pi}^2}{k^2} \bigg) I_1 \,,
\\
\textrm{Im} \, \tilde{\Gamma}_{\pi}(k^2) & = & - \frac{g^2}{8 \pi^2} \, m \,
\frac{k^2 - m^2 + m_\pi^2}{\lambda(k^2,m^2,m_{\pi}^2)} 
\Big( I_1 + (k^2 - m^2 - 2m_{\pi}^2) I_2 \Big) \,,
\end{eqnarray}
where we used 
$\lambda(k^2,m^2,m_{\pi}^2) = [k^2 - (m + m_{\pi})^2][k^2 - (m - m_{\pi})^2]$
and the integrals
\begin{eqnarray}
I_1 & = & \int d^4 l \, \delta(l^2 - m_{\pi}^2) \, \delta((k-l)^2 - m^2)
 \, = \, \frac{\pi}{2 k^2} \sqrt{\lambda(k^2,m^2,m_{\pi}^2)} \,
   \theta(k^2 - (m + m_{\pi})^2) \,,
\\
I_2 & = & \int d^4 l \, \frac{\delta(l^2 - m_{\pi}^2) \, \delta((k-l)^2 - m^2)}
                          {(k - P_h - l)^2 - m^2}
\nonumber \\
& = & - \frac{\pi}{2 \sqrt{\lambda(k^2,m^2,m_{\pi}^2)}}
   \ln \bigg(1 + \frac{\lambda(k^2,m^2,m_{\pi}^2)}
                      {k^2 m^2 - (m^2 - m_{\pi}^2)^2} \bigg) \,
   \theta(k^2 - (m + m_{\pi})^2) \,.
\end{eqnarray}
We note that the integral $I_2$ is evaluated for $P_h^2 = m_{\pi}^2$ and 
$(k-P_h)^2 = m^2$.

We are now in a position to check the ST sum rule~(\ref{e:sr2}), which in our
model takes the form
\begin{equation}
\int_0^1 dz \, z \, \Big( 2 \, m \, H_1^{\perp(1) \, q/q}(z) 
                         + m_{\pi} \, H_1^{\perp(1) \, \pi/q}(z) \Big) = 0 \,.  
\end{equation}
By making use of~(\ref{e:vertex_id}), one readily verifies that this sum rule is 
indeed satisfied if in either of the two results in~(\ref{e:coll_res1}) 
and~(\ref{e:coll_res2}) one makes the substitutions $z \to z' = 1 - z$ and 
$\vec{k}_T \to \vec{k}_T' = \frac{1-z}{z} \vec{k}_T$.
This means that neither the $z$-integration nor the $k_T$-integration has to be
performed explicitly.
One rather finds a cancellation of the contributions from~(\ref{e:coll_res1}) 
and~(\ref{e:coll_res2}) on the level of the integrand.

Finally, we want to explore if a result like the ST sum rule also holds for 
the three transverse momentum dependent FFs $D_{1T}^\perp$, $G_{1T}$, and 
$H_{1L}^\perp$ as was suggested in~\cite{Schafer:1999kn}.
In the model-independent part of our study we have only shown that the proof 
we gave for the ST sum rule does not apply to those FFs.
These three functions have in common that the final state hadron is polarized, 
which implies that in our model we only receive contributions from fragmentation 
into a quark.
We begin with the T-odd function $D_{1T}^\perp$, which is quite relevant for 
fragmentation into transversely polarized hyperons 
(see, e.g., Ref.~\cite{Anselmino:2000vs}).
It receives nonzero contributions from diagrams (e) and (f) in Fig.~\ref{f:2},
and the result reads
\begin{equation}
D_{1T}^{\perp \, q/q}(z,z^2 \vec{k}_T^2) = 
- H_{1}^{\perp \, q/q}(z,z^2 \vec{k}_T^2) \,,
\end{equation}
with $H_{1}^{\perp \, q/q}$ as given in Eq.~(\ref{e:coll_res1}).
One finds that for $D_{1T}^\perp$ a sum rule of the type~(\ref{e:sr2}) does not 
hold.
This can, for instance, be shown by focusing on the ultraviolet divergent part 
of the $k_T$-integral.
To be more specific, one finds
\begin{equation}
 \int_0^1 dz \, z \, D_{1T}^{\perp(1) \, q/q}(z) = 
 - \frac{g^4}{3 \times 2^{11} \, \pi^3} \, \ln^2 \frac{\Lambda^2}{m^2} 
 + \textrm{less singular} \,,
\end{equation}
where $\Lambda^2$ is an upper cutoff for the $k_T^2$-integration.
In contrast to the Collins function, for which fragmentation into a quark and 
fragmentation into a pion show up, for $D_{1T}^\perp$ the fragmentation into a quark 
is not compensated by another term.
The T-even functions $G_{1T}$ and $H_{1L}^\perp$ can be computed to ${\cal O}(g^2)$ 
on the basis of diagram (c) in Fig.~\ref{f:1} leading to
\begin{equation}
G_{1T}^{q/q}(z,z^2 \vec{k}_T^{\,2}) = 
H_{1L}^{\perp \, q/q}(z,z^2 \vec{k}_T^{\,2}) =
\frac{g^2}{16\pi^3} \,
\frac{(1-z)^2 \, m^2}
     {z^3 \, \Big( \vec{k}_T^{\,2} + \frac{(1-z)^2}{z^2}m^2 
                 + \frac{m_{\pi}^2}{z} \Big)^2} \,.
\end{equation}
Again, by just focusing on the ultraviolet divergent part of the $k_T$-integral,
one also readily verifies that for these two FFs a sum rule of the 
type~(\ref{e:sr2}) cannot exist.
Explicit calculation provides
\begin{equation}
 \int_0^1 dz \, z \, G_{1T}^{(1) \, q/q}(z) = 
 \int_0^1 dz \, z \, H_{1L}^{\perp(1) \, q/q}(z) = 
 \frac{g^2}{96 \pi^2} \, \ln \frac{\Lambda^2}{m^2} + \textrm{ultraviolet finite} \,.
\end{equation}

%
%
\section{Summary}
\noindent
In this note, momentum sum rules for fragmentation functions have been studied 
by performing both a model-independent analysis as well as explicit model 
calculations.
In particular, we have provided a general field-theoretical proof of the 
ST sum rule~\cite{Schafer:1999kn} for the Collins function 
$H_1^\perp$~\cite{Collins:1992kk} in QCD.
The existing derivation of the ST sum rule was merely based on intuitive 
arguments about conservation of transverse momentum in the fragmentation 
process~\cite{Schafer:1999kn}.
In this respect, there is a strong similarity between the ST sum rule and the 
longitudinal momentum sum rule for the unpolarized fragmentation function 
$D_1$:~they are both intuitive, but their general proof in QCD is more 
involved~\cite{Collins:1981uw}.
The same statement also applies to the so-called Burkardt sum 
rule~\cite{Burkardt:2003yg,Burkardt:2004ur} for the transverse momentum dependent 
Sivers parton distribution~\cite{Sivers:1989cc,Sivers:1990fh}.

In the literature it was suggested that the ST sum rule should also hold for 
the three additional transverse momentum dependent 
FFs $D_{1T}^\perp$, $G_{1T}$, and 
$H_{1L}^\perp$~\cite{Schafer:1999kn,Boer:1999ya,Anselmino:2000vs}.
However, here we have shown that the general proof of the ST sum rule cannot be 
extended to these cases.
We have also demonstrated that, in the light-cone gauge, the proof of the 
longitudinal momentum sum rule for $D_1$ is not spoiled by the relatively
recently discovered transversely running Wilson lines in the fragmentation
correlator.

We have exploited a simple self-consistent quark-pion coupling model in order to 
explicitly verify/falsify the momentum sum rules.
Though the model does not know about all the complexities of QCD, it nevertheless 
can be used for interesting cross checks.
We have been able to verify the sum rule for $D_1$ as well as the ST sum rule for 
the Collins function $H_1^\perp$ to lowest nontrivial order in the coupling 
constant.
On the other hand, we have shown explicitly that the ST sum rule does not hold
for the aforementioned additional three FFs.
\\[0.5cm]
%
%
\noindent
{\bf Acknowledgments:} 
The work is partially supported by the Verbundforschung 
``Hadronen und Kerne'' of the BMBF. 
A.M. acknowledges the support of the NSF under Grant No. PHY-0855501.

%
%

\end{document}